\documentclass[12pt]{article}
\usepackage{epsf}
\title{ {\bf Unparticle physics and lepton flavor violating radion
decays in the Randall-Sundrum scenario}}
\author{\vspace{1cm}\\
        {\bf E. O. Iltan,}
        \thanks{E-mail address:
        eiltan@newton.physics.metu.edu.tr}
 \\
        Physics Department, Middle East Technical University \\
        Ankara, Turkey\\}

\date{}

\begin{document}
\setlength{\baselineskip}{24pt}
\maketitle
\setlength{\baselineskip}{7mm}
\begin{abstract}
We predict the branching ratios of the lepton flavor violating
radion decays $r\rightarrow e^{\pm} \mu^{\pm}$, $r\rightarrow
e^{\pm} \tau^{\pm}$ and $r\rightarrow \mu^{\pm} \tau^{\pm}$  in
the framework of the Randall-Sundrum scenario that the lepton
flavor violation is carried by the scalar unparticle mediation. We
observe that their BRs are strongly sensitive to the unparticle
scaling dimension and, for its small values,  the  branching
ratios can reach to the values of the order of $10^{-8}$, for the
heavy lepton flavor case.
\end{abstract}
\thispagestyle{empty}
\newpage
\setcounter{page}{1}
\section{Introduction}
Recently, Georgi \cite{Georgi1,Georgi2} proposed unparticle stuff
which has non-integer scaling dimension $d_u$ and looks like a
number of $d_u$ massless invisible particles. The idea behind is
the existence of a non-trivial scale invariant sector beyond the
standard model (SM) , with non-trivial infrared fixed point and
the scaling dimension $d_u$. Georgi suggested that, at the energy
scales around $\Lambda_U\sim 1\,TeV$, this sector appears as so
called unparticle stuff. The effective lagrangian can be
constructed to describe the interactions of unparticles with the
SM fields in the low energy level and this approach opens a window
to test the effects of the possible scale invariant sector,
experimentally.

The missing energies at various processes which can be measured at
LHC or $e^+e^-$ colliders, the dipole moments of fundamental
particles and the processes in which the unparticle(s) appears as
mediator are the possible candidates in  order to search the
effects of unparticle(s). There is various phenomenological work
done in the literature on unparticles
\cite{Georgi2}-\cite{TKikuchi}: their effects on the missing
energy of many processes, the anomalous magnetic moments, the
electric dipole moments, $D^0-\bar{D}^0$ and $B^0-\bar{B}^0$
mixing, lepton flavor violating interactions, direct CP violation
in particle physics; the phenomenological implications in
cosmology and in astrophysics.

In this work,  we study the lepton flavor violating (LFV) decays
of the Randall Sundrum (RS1) radion field in the case that the LF
violation\footnote{In the SM with massive neutrinos, so called
$\nu$SM \cite{Pontecorvo1}-\cite{Pontecorvo3}, the lepton mixing
mechanism is permitted. However, the negligibly small branching
ratios (Brs) of LFV decays stimulate one to search for new LF
violation mechanisms. In the present work, we do not take into
account the LF violation coming from the possible massive
neutrinos.} is carried by the scalar unparticle mediation. The RS1
model is based on the non-factorizable geometry \cite{Rs1,Rs2} in
the five space-time dimensions and the extra dimension is
compactified into $S^1/Z_2$ orbifold with two 4D brane boundaries.
In one of the boundary, the so called Planck brane, the gravity is
localized and in the another one, the TeV brane, all other fields
are restricted. The size of extra dimension is proportional to the
vacuum expectation of a scalar field and its fluctuation over the
expectation value is called the radion field that has been studied
extensively in the literature
\cite{Goldberger}-\cite{IltanKorutlu}. Here, we predict the BRs of
the LFV decays in the framework of the RS1 scenario, by using the
effective lagrangian in order to insert the possible scalar
unparticle mediation. We observe that the BRs of the processes we
study are strongly  sensitive to the unparticle scaling dimension
$d_u$ and, for its small values $d_u < 1.1$, the BRs enhance
considerably.

The paper is organized as follows: In Section 2, we present the
the effective lagrangian and effective vertices which drive the
LFV  decays with scalar unparticle mediation, by respecting the
RS1 scenario. Furthermore, we give the expression for their BRs.
Section 3 is devoted to the discussion and to our conclusions. In
the appendix, we present the interaction vertices including the
radion field.
\section{The LFV RS1 radion decay with scalar unparticle mediation}
The LFV processes are among the rare decays in the sense that they
exist at least in the one loop level and their BRs are small.
However, their existence in the loop level makes them worthwhile
to analyze since physical quantities related to them contain
considerable information about the model used and the free
parameters existing. In the present work, we study the LFV decays
of the radion field, in the framework of the RS1 scenario. Here we
assume that the LF violation
is carried by the unparticle stuff which is introduced by Georgi
\cite{Georgi1, Georgi2}. The starting point of the idea is the
interaction of two sectors, the SM and the ultraviolet sector with
non-trivial infrared fixed point, at high energy level. The
ultraviolet sector appears as new degrees of freedom, called
unparticles, being massless and having non integral scaling
dimension $d_u$ around, $\Lambda_U\sim 1\,TeV$. This mechanism
results in the existence of the  effective field theory with
effective Lagrangian in the low energy level and the corresponding
Lagrangian reads
\begin{equation}
{\cal{L}}_{eff}\sim
\frac{\eta}{\Lambda_U^{d_u+d_{SM}-n}}\,O_{SM}\, O_{U} \,,
\label{efflag}
\end{equation}
where $O_U$ is the unparticle operator, the parameter $\eta$ is
related to the energy scale of ultraviolet sector, the low energy
one and the matching coefficient \cite{Georgi1,Georgi2,Zwicky} and
$n$ is the space-time dimension.

At this stage, we would like to give a brief explanation about the
RS1 scenario and the effective Lagrangian which is responsible for
the LFV decay underconsideration. The RS1 model is formulated in
the warped extra dimension, which is compactified into $S^1/Z_2$
orbifold. There exist two 4D surfaces (branes), which are
boundaries of the extra dimension, in 5D world. One of the brane
is called the Planck brane where the gravity, extending into the
bulk with varying strength, peaks near and the other one is called
the TeV brane where we live. The considered behavior of the
gravity results in the explanation of the well known hierarchy
problem. In addition to this, the cosmological constant problem is
solved with the help of the equal and opposite tensions in these
two branes\footnote{The 5D cosmological constant does not vanish,
however, the low energy effective theory has flat 4D space-time by
considering  both branes have equal and opposite tensions.}. The
background metric of this 5D world is
\begin{eqnarray}
ds^2=e^{-2\,A(y)}\,\eta_{\mu\nu}\,dx^\mu\,dx^\nu-dy^2\, ,
\label{metric1}
\end{eqnarray}
where $A(y)=k\,|y|$, $k$ is the bulk curvature constant, $y$ is
the extra dimension parametrized as $y=R\,\theta$. The exponential
factor $e^{-k\,L}$ with the inter brane separation $L=R\,\pi$, is
the warp factor which ensures that all the mass terms are rescaled
in the TeV brane. With a rough estimate  $L\sim 30/k$ the
hierarchy problem is solved and all mass terms are pulled down to
the TeV scale. In this scenario, the radion field appears as
fluctuation over the expectation value of the field $L(x)$ that
its vacuum expectation value is related to the size $L$ of extra
dimension. On the other hand, the field $L(x)$ should acquire a
mass not to have a conflict with the equivalence principle and,
the radion field can be stabilized with a mechanism, proposed by
Goldberger and Wise \cite{Goldberger}. Including the radial
fluctuations, the metric in 5D is defined as \cite{Rubakov}
\begin{eqnarray}
ds^2=e^{-2\,A(y)-2\,F(x)}\,\eta_{\mu\nu}\,dx^\mu\,dx^\nu- (1+2\,
F(x)\,dy^2\, , \label{metric2}
\end{eqnarray}
where $F(x)$ the scalar field,
\begin{eqnarray}
F(x)=\frac{1}{\sqrt{6}\,M_{Pl}\,e^{-k\,L}}\, r(x)\, , \label{Fx}
\end{eqnarray}
and $r(x)$ is the normalized radion field (see for example
\cite{CsabaMichael}). Finally, the the induced metric at the
orbifold point $\theta=\pi$ (TeV brane) reads,
\begin{eqnarray}
g^{ind}_{\mu\nu}=e^{-2\,A(L)-2\frac{\gamma}{v}\,r(x)}\,\eta_{\mu\nu}
\, , \label{metricind}
\end{eqnarray}
with $\gamma=\frac{v}{\sqrt{6}\,\Lambda_r}$,
$\Lambda_r=M_{Pl}\,e^{-k\,L}$ and $v$ is the vacuum expectation
value of the SM Higgs boson.

Here, we are ready the construct the effective interaction
lagrangian for the LFV decays we study.
Notice that we choose the appropriate operators with lowest
possible dimension since they have the most powerful effect in the
low energy effective theory (see for example \cite{SChen}). The
part of the effective lagrangian which is responsible for the LF
violation reads
\begin{eqnarray}
{\cal{L}}_1= \frac{\sqrt{-g^{ind}}}{\Lambda^{du-1}}\Big
(\lambda_{ij}^{S}\, \bar{l}_{i}
\,l_{j}+\lambda_{ij}^{P}\,\bar{l}_{i} \,i\,\gamma_5\,l_{j}\Big)\,
O_{U} \, , \label{lagrangianscalar}
\end{eqnarray}
where $l$ is the lepton field,  $O_{U}$ is the scalar unparticle
(\textit{U}) operator and $\lambda_{ij}^{S}$ ($\lambda_{ij}^{P}$)
is the scalar (pseudoscalar)  coupling. In addition to the tree
level $\textit{U}-l_1-l_2$ interaction with the coupling
$\sim\,\frac{1}{\Lambda^{du-1}}\Big(\lambda_{ij}^{S}+i\,
\lambda_{ij}^{P}\,\gamma_5\Big)$, the lagrangian in eq.
(\ref{lagrangianscalar}) drives the four point
$r-\textit{U}-l_1-l_2$ interaction due to the factor
$\sqrt{-g^{ind}}=e^{-4\,A(L)-4\frac{\gamma}{v}\,r(x)}$ (see Fig.
1-a). Here the background term $e^{-4\,A(L)}$ in $\sqrt{-g^{ind}}$
is embedded into the redefinitions of the fields on the TeV brane,
namely, they are warped as  $l\rightarrow
e^{3\,A(L)/2}\,l_{warp}$, $\frac{O_{U}}{\Lambda^{du-1}}\rightarrow
e^{A(L)}\,(\frac{O_{U}}{\Lambda^{du-1}})_{warp}$, and in the
following we use warped fields without the $warp$ index. Since the
FV radion decays, $r\rightarrow l_1^- l_2^+$, can exist at least
in one loop level (see Fig.\ref{figselfvert}), one needs the
scalar unparticle propagator which is obtained by using the scale
invariance. The two point function of the unparticle results in
\cite{Georgi2, Cheung1}
\begin{eqnarray}
\int\,d^4x\,
e^{ipx}\,<0|T\Big(O_U(x)\,O_U(0)\Big)0>=i\frac{A_{d_u}}{2\,\pi}\,
\int_0^{\infty}\,ds\,\frac{s^{d_u-2}}{p^2-s+i\epsilon}=i\,\frac{A_{d_u}}
{2\,sin\,(d_u\pi)}\,(-p^2-i\epsilon)^{d_u-2} \, ,
\label{propagator}
\end{eqnarray}
with the factor $A_{d_u}$
\begin{eqnarray}
A_{d_u}=\frac{16\,\pi^{5/2}}{(2\,\pi)^{2\,d_u}}\,\frac{\Gamma(d_u+\frac{1}{2})}
{\Gamma(d_u-1)\,\Gamma(2\,d_u)} \, . \label{Adu}
\end{eqnarray}
The function $\frac{1}{(-p^2-i\epsilon)^{2-d_u}}$ in eq.
(\ref{propagator}) becomes
\begin{eqnarray}
\frac{1}{(-p^2-i\epsilon)^{2-d_u}}\rightarrow
\frac{e^{-i\,d_u\,\pi}}{(p^2)^{2-d_u}} \, , \label{strongphase}
\end{eqnarray}
for $p^2>0$ and a non-trivial phase appears as a result of
non-integral scaling dimension.

On the other hand, the part of the lagrangian which carries the
interaction of leptons with the radion field reads (for example
see \cite{PKDas1,PKDas2})
\begin{eqnarray}
{\cal{L}}_{2}&=& \sqrt{-g^{ind}}\, \Bigg( g^{ind\,\, \mu\nu}\,
\bar{l}\,\gamma_\mu\,i\,D_\nu\,l-m_l\,\bar{l}\,l\Bigg) \, ,
\label{lagrangian2}
\end{eqnarray}
where
\begin{eqnarray}
D_\mu \,l=\partial_\mu\,l+\frac{1}{2}\,w_\mu^{ab}\,\Sigma_{ab}\,l
\, , \label{Dmuf}
\end{eqnarray}
and $\Sigma_{ab}=\frac{1}{4}[\gamma_a,\gamma_b]$. Here
$w_\mu^{ab}$ is the spin connection which reads
\begin{eqnarray}
w_\mu^{ab}=-\frac{\gamma}{v}\partial_\nu\,r\,(e^{\nu
b}\,e^a_\mu-e^{\nu a}\,e_\mu^b )\, , \label{vierbein}
\end{eqnarray}
linear in $r$. In this equation, $e^a_\mu$ are the vierbein fields
and they satisfy the relation
\begin{eqnarray}
e^\mu_a\,e^{a \nu}=g^{ind\,\, \mu\nu}\, . \label{metrvierbein}
\end{eqnarray}
Using eqs. (\ref{lagrangian2})-(\ref{metrvierbein}), one gets the
part of the lagrangian which drives the tree level $l-l-r$
interaction (see Fig.1-b) as
\begin{eqnarray}
{\cal{L'}}_{2}&=&\Bigg\{ -3\frac{\gamma}{v}\, r\,
\bar{l}\,i\,\partial\!\!\!/ l-3\frac{\gamma}{2\,v}\,
\bar{l}\,i\,\partial\!\!\! / r\, l+ 4\,\frac{\gamma}{v}\, m_l\,r\,
\bar{l}\,l \Bigg\} \, . \label{lagrangian2p}
\end{eqnarray}

Now, we present the matrix element square of the LFV radion decay
which exits at least in one loop order (see Fig. \ref{figselfvert}
for the possible vertex and self energy diagrams):
\begin{eqnarray}
|M|^2= 2\Big( m_r^2 -(m_{l_1^-}+m_{l_2^+})^2\Big)\,|A|^2+2\Big(
m_r^2 -(m_{l_1^-}-m_{l_2^+})^2\Big)\,|A'|^2 \, , \label{Matrx2}
\end{eqnarray}
where
\begin{eqnarray}
A&=&\int^{1}_{0}\,dx\,f_{self}^S+\int^{1}_{0}\,dx\,\int^{1-x}_{0}\,dy\,
f_{vert}^S \, ,\nonumber \\
A'&=&\int^{1}_{0}\,dx\,f_{self}^{\prime\,S}+\int^{1}_{0}\,dx\,
\int^{1-x}_{0}\,dy\, f_{vert}^{\prime\,S} \nonumber \, , \\
\label{funpart}
\end{eqnarray}
and the explicit expressions of $f_{self}^S$,
$f_{self}^{\prime\,S}$, $f_{vert}^S$, $f_{vert}^{\prime\,S}$ read
\begin{eqnarray}
f_{self}^S&=& \frac{i\,c_1\,(1-x)^{1-d_u}}{32\,\pi^2\,\Big(
m_{l_2^+}-m_{l_1^-}\Big)\,(1-d_u)}\,\sum_{i=1}^3\,
 \Big\{(\lambda_{il_1}^S\,
\lambda_{il_2}^S+\lambda_{il_1}^P \lambda_{il_2}^P)
\,(1-x)\nonumber \\ &\times& \Big(
m_{l_1^-}\,(5\,m_{l_1^-}-3\,m_{l_2^+})\,
L_{self}^{d_u-1}-m_{l_2^+}\, (5\,m_{l_2^+}-3\,m_{l_1^-})\,
L_{self}^{\prime d_u-1} \Big) \nonumber
\\ &+&
(\lambda_{il_1}^P\, \lambda_{il_2}^P-\lambda_{il_1}^S
\lambda_{il_2}^S) \,m_i\,\Big( (3\,m_{l_2^+}-5\,m_{l_1^-})\,
L_{self}^{d_u-1}-(3\,m_{l_1^-}-5\,m_{l_2^+})\,L_{self}^{\prime
d_u-1} \Big)\Big\} \, , \nonumber \\ \nonumber \\
f_{self}^{\prime\,S}&=&
\frac{c_1\,(1-x)^{1-d_u}]}{32\,\pi^2\,\Big(
m_{l_2^+}+m_{l_1^-}\Big)\,(1-d_u)}\,\sum_{i=1}^3\,
 \Big\{(\lambda_{il_1}^S\,
\lambda_{il_2}^P-\lambda_{il_1}^P \lambda_{il_2}^S)
\,(1-x)\nonumber \\ &\times& \Big(
m_{l_1^-}\,(3\,m_{l_2^+}+5\,m_{l_1^-})\,
L_{self}^{d_u-1}-m_{l_2^+}\,(3\,m_{l_1^-}+5\,m_{l_2^+})\,L_{self}^{\prime
d_u-1}  \Big) \nonumber
\\ &+&
(\lambda_{il_1}^S\, \lambda_{il_2}^P+\lambda_{il_1}^P
\lambda_{il_2}^S) \, m_i\,\Big( (3\,m_{l_2^+}+5\,m_{l_1^-})\,
L_{self}^{d_u-1}+(3\,m_{l_1^-}+5\,m_{l_2^+})\,L_{self}^{\prime
d_u-1}) \Big)\Big\} \, , \nonumber \\ \nonumber \\
f_{vert}^{S}&=& \frac{-3\,i\,c_1\,(1-x-y)^{1-d_u}}{32\,\pi^2}\,
\sum_{i=1}^3\,\frac{1}{\,L_{vert}^{2-d_u}}\,
 \Bigg\{(\lambda_{il_1}^S\,
\lambda_{il_2}^S+\lambda_{il_1}^P \,\lambda_{il_2}^P)\,
\Big\{(1-x-y)\nonumber \\&\times&\Bigg( m_{l_2^+}\,m_{l_1^-}
(m_{l_2^+}+m_{l_1^-})\,(1-x-y+2\,x\,y)+2\,\Big(
m_{l_2^+}^3\,y\,(y-1)+m_{l_1^-}^3\,x\,(x-1)\Big)\Bigg) \nonumber \\
&-& \frac{3\,L_{vert}}{1-d_u}\, \Big(
m_{l_1^-}\,(2\,x-1)+m_{l_2^+}\,(2\,y-1) \Big)\nonumber \\ &+&
m_r^2\,\Big(m_{l_1^-}\,x\,(2\,y-1)\,(x-1)+m_{l_2^+}\,y\,(2\,x-1)\,(y-1)\Big)
\Bigg) \nonumber \\ &+& 5\,m_i^2\,\Big(
m_{l_1^-}\,(2\,x-1)+m_{l_2^+}\,(2\,y-1)\Big) \Big\}\nonumber \\
&+& (\lambda_{il_1}^P\,\lambda_{il_2}^P-\lambda_{il_1}^S
\lambda_{il_2}^S)\, \Big\{ (1-x-y)\,m_i\,\Big(2\,
m_{l_1^-}\,m_{l_2^+}+m_{l_1^-}^2\,(4\,x-3)+m_{l_2^+}^2\,(4\,y-3)\Big)\nonumber
\\ &-& \frac{8\,m_i}{1-d_u}
\,L_{vert}-m_i\,m_r^2\,(3\,x+3\,y-4\,x\,y)+8\,m_i^3 \Big\} \Bigg\}
\, , \nonumber \\ \nonumber \\ \nonumber \\
f_{vert}^{\prime\,S}&=&
\frac{-3\,c_1\,(1-x-y)^{1-d_u}}{32\,\pi^2}\,\sum_{i=1}^3\,
\frac{1}{\,L_{vert}^{2-d_u}}\,
 \Bigg\{(\lambda_{il_1}^P \,\lambda_{il_2}^S-\lambda_{il_1}^S\,
\lambda_{il_2}^P)\, \Big\{(1-x-y)\nonumber \\
&\times& \Bigg( m_{l_2^+}\,m_{l_1^-}\,(m_{l_2^+}-m_{l_1^-})
(1-x-y+2\,x\,y)+2\,\Big(
-m_{l_2^+}^3\,y\,(y-1)+m_{l_1^-}^3\,x\,(x-1)\Big)\Bigg) \nonumber \\
&-& \frac{3\,L_{vert}}{1-d_u}\, \Big(
m_{l_1^-}\,(2\,x-1)-m_{l_2^+}\,(2\,y-1) \Big)\nonumber \\ &+&
m_r^2\,\Big(m_{l_1^-}\,x\,(2\,y-1)\,(x-1)-m_{l_2^+}\,y\,(2\,x-1)\,(y-1)\Big)
\Bigg) \nonumber \\ &+& 5\,m_i^2\,\Big(
m_{l_1^-}\,(2\,x-1)-m_{l_2^+}\,(2\,y-1)\Big) \Big\}\nonumber \\
&+& (\lambda_{il_1}^S\,\lambda_{il_2}^P+\lambda_{il_1}^P
\lambda_{il_2}^S)\, \Big\{ (1-x-y)\,m_i\,\Big(-2\,
m_{l_1^-}\,m_{l_2^+}+m_{l_1^-}^2\,(4\,x-3)+m_{l_2^+}^2\,(4\,y-3)\Big)\nonumber
\\ &-& \frac{8\,m_i}{1-d_u}
\,L_{vert}-m_i\,m_r^2\,(3\,x+3\,y-4\,x\,y)+8\,m_i^3 \Big\} \Bigg\}
\, , \label{spcouplings}
\end{eqnarray}
with
\begin{eqnarray}
L_{self}&=&x\,\Big(m_{l_1^-}^2\,(1-x)-m_i^2\Big)
\, , \nonumber \\
L_{self}^{\prime}&=&x\,\Big(m_{l_2^+}^2\,(x-1)-m_i^2\Big) \, ,
\nonumber \\
L_{vert}&=&(m_{l_1^-}^2\,x+m_{l_2^+}^2\,y)\,(1-x-y)-m_i^2\,(x+y)+m_r^2\,x\,y
\, ,
\end{eqnarray}
and
\begin{eqnarray}
c_1&=&\frac{\gamma\,A_{d_u}}{2\,v\,sin\,(d_u\pi)\,\Lambda_u^{2\,(d_u-1)}}\,
.
\end{eqnarray}
In eq. (\ref{spcouplings}), the flavor changing scalar and
pseudoscalar couplings $\lambda_{il_{1(2)}}^{S,P}$ represent the
effective interaction between the internal lepton $i$,
($i=e,\mu,\tau$) and the outgoing $l_1^-\,(l_2^+)$ lepton (anti
lepton).

Finally, the BR for $r\rightarrow l_1^-\,l_2^+$ can be obtained by
using the matrix element square as
\begin{eqnarray}
BR (r\rightarrow l_1^- \,l_2^+)=\frac{1}{16\,\pi\,m_r}\,
\frac{|M|^2}{\Gamma_r}\, , \label{BR1}
\end{eqnarray}
with the radion total decay width $\Gamma_r$. In the numerical
analysis,  we consider the BR due to the production of sum of
charged states, namely,
\begin{eqnarray}
BR (r\rightarrow l_1^{\pm}\,l_2^{\pm})= \frac{\Gamma(r\rightarrow
(\bar{l}_1\,l_2+\bar{l}_2\,l_1))}{\Gamma_r} \, .\label{BR2}
\end{eqnarray}
%
\section{Discussion}
In this section, we analyze the LFV  RS1 radion decays
$r\rightarrow l_1^- l_2^+$ in the case that the flavor violation
is carried by the scalar unparticle mediation and we estimate
their BRs. The LFV  radion decays $r\rightarrow l_1^- l_2^+$ can
exist at least in one loop level and, in the present case, the
flavor violation is driven by the fermion-fermion-\textit{U}
couplings, in the effective theory. The scaling dimension of the
unparticle, the couplings, the radion mass and the energy scale is
the free parameters of the scenario studied. At this stage, we
discuss our restrictions for these free parameters:
\begin{itemize}
\item The scaling dimension $d_u$ is chosen  in the range $1< d_u
<2$. For $d_u>1$, one is free from the non-integrable singularity
problem in the decay rate \cite{Georgi2}. On the other hand the
momentum integrals converges for $d_u<2$ \cite{Liao1}.
\item  Here we consider that the diagonal couplings $\lambda_{ii}$
respects the lepton family the hierarchy,
$\lambda_{\tau\tau}>\lambda_{\mu\mu}>\lambda_{ee}$, and the
off-diagonal couplings, $\lambda_{ij}, i\neq j$ are family blind
and universal. Furthermore, we take the off diagonal couplings as,
$\lambda_{ij}=\kappa \lambda_{ee}$ with $\kappa < 1$. In our
numerical calculations, we choose $\kappa=0.5$ and take the
greatest numerical value of diagonal couplings not more than the
order of one.
\item For the mass of the radion, we choose the values
$m_r=200\,(GeV)$, $m_r=500\,(GeV)$ and $m_r=800\,(GeV)$ to observe
the radion mass dependence of the BRs of the decays under
consideration.
\item  We take the energy scale at least $\Lambda_u=10\,(TeV)$ and
study $\Lambda_u$ dependence of the BRs for its various values.
\end{itemize}
For the calculation of the BRs of $r\rightarrow l_1^- l_2^+$
decays we need the total decay width $\Gamma_r$ of the radion and
we use the theoretical predictions given in the literature. The
decay width is dominated by $r\rightarrow gg$ for the radion mass
$m_r\le 150\,GeV$. This quantity is calculated from the trace
anomaly which exists in one loop order. For the higher masses of
the radion, which are beyond the WW and ZZ thresholds, the main
decay mode is $r\rightarrow WW$. In this work, we take the total
decay width $\Gamma_r$ of the radion by considering the dominant
decays $r\rightarrow gg\, (\gamma\gamma, ff, W^+W^-,ZZ, SS)$ where
$S$ are the neutral Higgs particles (see \cite{Cheung} for the
explicit expressions of these decay widths). Notice that we
include the possible processes in the $\Gamma_r$ according to the
mass of the radion, in our calculations.
Throughout our calculations we use the input values given in Table
(\ref{input}).
\begin{table}[h]
        \begin{center}
        \begin{tabular}{|l|l|}
        \hline
        \multicolumn{1}{|c|}{Parameter} &
                \multicolumn{1}{|c|}{Value}     \\
        \hline \hline
        $m_e$           & $0.0005$   (GeV)  \\
        $m_{\mu}$                   & $0.106$ (GeV) \\
        $m_{\tau}$                  & $1.780$ (GeV) \\
        \hline
        \end{tabular}
        \end{center}
\caption{The values of the input parameters used in the numerical
          calculations.}
\label{input}
\end{table}
\newpage
In  Fig.\ref{rmuedu}, we present the contribution of the scalar
unparticle to the BR $(r\rightarrow \mu^{\pm}\, e^{\pm})$ with
respect to the scale parameter $d_u$, for the energy scale
$\Lambda_u=10\, TeV$, the couplings $\lambda_{ee}=0.01$ and
$\lambda_{\mu\mu}=0.1$. Here the solid (dashed, small dashed) line
represents the BR  for $m_r=200\,(GeV)$ ($m_r=500\,(GeV),
m_r=800\,(GeV)$). The BR of the decay considered is strongly
sensitive to the scale $d_u$ and, with the decreasing values of
$d_u$, there is a considerable enhancement in the BR. The BR
reaches to the numerical values $10^{-10}$, for $d_u<1.1$ and
$m_r=200\,(GeV)$. For the heavier mass values of the radion the BR
is suppressed and it decreases to the values of the order of
$10^{-12}$ for $d_u<1.1$ and $m_r=500,\, 800\,(GeV)$.

In  Fig.\ref{rtauedu}, we show the contribution of the scalar
unparticle to the BR $(r\rightarrow \tau^{\pm}\, e^{\pm})$ with
respect to the scale parameter $d_u$, for the energy scale
$\Lambda_u=10\, TeV$, the couplings $\lambda_{ee}=0.01$ and
$\lambda_{\tau\tau}=1$. Here the solid (dashed, small dashed) line
represents the BR  for $m_r=200\,(GeV)$ ($m_r=500\,(GeV),
m_r=800\,(GeV)$). For $d_u<1.1$ the BR increases considerably and
it is in the range $10^{-10}-10^{-6}$, for $m_r=200\,(GeV)$. For
the radion masses $m_r=500,\, 800\,(GeV)$, the BR is suppressed to
the values $10^{-12}$ for $d_u \sim 1.1$.

Fig.\ref{rtaumudu} represents the contribution of the scalar
unparticle to the BR $(r\rightarrow \tau^{\pm}\, \mu^{\pm})$ with
respect to the scale parameter $d_u$, for the energy scale
$\Lambda_u=10\, TeV$, the couplings $\lambda_{\mu\mu}=0.1$ and
$\lambda_{\tau\tau}=1$. Here the solid (dashed, small dashed) line
represents the BR for $m_r=200\,(GeV)$ ($m_r=500\,(GeV),
m_r=800\,(GeV)$). The BR for $d_u<1.1$ and $m_r=200\,(GeV)$ is
large similar to the $(r\rightarrow \tau^{\pm}\, e^{\pm})$ decay,
and it is in the range of $10^{-10}-10^{-6}$, in this region of
the scale dimension.

These figures show that the BRs of the LFV decays are sensitive to
the scaling dimension $d_u$ and they become negligibly small
(quite large) for the values of the scaling dimension $d_u$
greater and far from $1.1$ (near to $1.0$).

Now, we would like to analyze the energy scale $\Lambda_u$ and the
parameter $\lambda$ dependence of the BRs of the LFV decays in the
various figures, for completeness.

Fig. \ref{rmueLamu} (\ref{rtaueLamu}, \ref{rtaumuLamu}) is devoted
to the contribution of the scalar unparticle to the BR
$(r\rightarrow \mu^{\pm}\, e^{\pm})$ (BR $(r\rightarrow
\tau^{\pm}\, e^{\pm})$, BR $(r\rightarrow \tau^{\pm}\,
\mu^{\pm})$) with respect to the energy scale $\Lambda_u$, for
$m_r=200\,(GeV)$ and the couplings $\lambda_{ee}=0.01$ and
$\lambda_{\mu\mu}=0.1$ ($\lambda_{ee}=0.01$ and
$\lambda_{\tau\tau}=1.0$, $\lambda_{\mu\mu}=0.1$ and
$\lambda_{\tau\tau}=1.0$). Here the solid-dashed-small dashed
lines represent the BR for $d_u=1.1-d_u=1.2-d_u=1.5$. The
increasing values of the energy scale $\Lambda_u$ result in the
suppression in the BR and the numerical values of the order of
$10^{-13}$ ($0.5\,10^{-10}$, 0.5\,$10^{-10}$) can be reached for
the energy scale $\Lambda_u$, $\Lambda_u <10\,(TeV)$ and
$d_u=1.1$. This figure shows also the scale $d_u$ sensitivity of
the BRs of the decays under consideration.

Fig. \ref{rmuecoupl} (\ref{rtauecoupl}, \ref{rtaumucoupl})
represents the the contribution of the scalar unparticle to the BR
$(r\rightarrow \mu^{\pm}\, e^{\pm})$ (BR $(r\rightarrow
\tau^{\pm}\, e^{\pm})$, BR $(r\rightarrow \tau^{\pm}\,
\mu^{\pm})$) with respect to the parameter $\lambda$, for the
energy scale $\Lambda_u=10\, TeV$ and the radion mass
$m_r=200\,(GeV)$. Here, we consider that the parameter $\lambda$
is connected to the couplings by the equalities,
$\lambda_{ee}=\lambda$, $\lambda_{\mu\mu}=10\,\lambda$,
$\lambda_{\tau\tau}=100\,\lambda$. The  solid-dashed-small dashed
lines represent the BR for $d_u=1.1-d_u=1.2-d_u=1.5$. It is
observed that the BR is strongly sensitive to the parameter
$\lambda$ as expected and, its numerical value is of the order of
$10^{-13}$ ($0.5\,10^{-10}$, 0.5\,$10^{-10}$) for $\lambda \sim
0.01$, for the scaling dimension $d_u$=1.1.

As a summary, the LFV decays of the radion in the RS1 model are
strongly  sensitive to the unparticle scaling dimension and, for
its small values $d_u < 1.1$, the BRs enhance considerably. The
other free parameters of the scenario studied are the
$\textit{U}$-fermion-fermion couplings, the energy scale and the
radion mass and the dependencies of the BRs of LFV decays to these
free parameters are also strong. The possible production of the
radion (the most probable production is due to the gluon fusion,
$gg\rightarrow r$ \cite{Cheung}) would stimulate one to study its
LFV decays and the forthcoming experimental results would be
instructive in order to test the possible signals coming from the
extra dimensions and the new physics which drives the flavor
violation, here is the unparticle physics.
\vspace{2cm}
\appendix
\section{The vertices including the radion field, in the present work}
In this section we present the vertices, including the radion
field, used in our calculations.
\begin{figure}
\vspace{3cm}
\begin{tabular}{p{6cm} p{5cm}} \vskip -7.6truein
\parbox[b]{6cm}{\epsffile{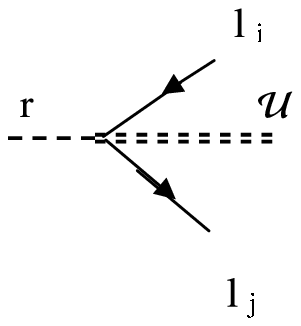}} & \vspace{0.1cm}\\ \vskip -0.4truein
\hspace{4.7cm}
\raisebox{5.ex}{$\,\,\,\,\,\,\,\frac{4\,i\,\gamma}{v\,\Lambda^{du-1}}
\left[\lambda_{ij}^{S}+i\,\lambda_{ij}^{P}\,\gamma_5\right]$}
%
%
\textbf{(a)}\\ \\ \\ \vskip -7.3truein
\parbox[b]{16cm}{\epsffile{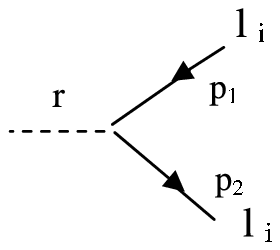}} &\vspace{0.2cm}\\ \hspace{4.8cm}
\raisebox{5.ex}{$\frac{-i\,\gamma}{v}\,\left[\frac{3}{2}\,
(p_1\!\!\!\!\!/+p_2\!\!\!\!\!/\,)-4\,m_{l_i}\right]$}
\\
\textbf{(b)}\\ \\ \vskip -7.1truein
\\ \\ \\ \\
\end{tabular}
%
\label{figvert1} \caption{The vertices including the radion
field.}
\end{figure}
%
%
%
\newpage
\newpage
\begin{figure}[htb]
\vskip -7.2truein \centering \epsfxsize=9.5in
\leavevmode\epsffile{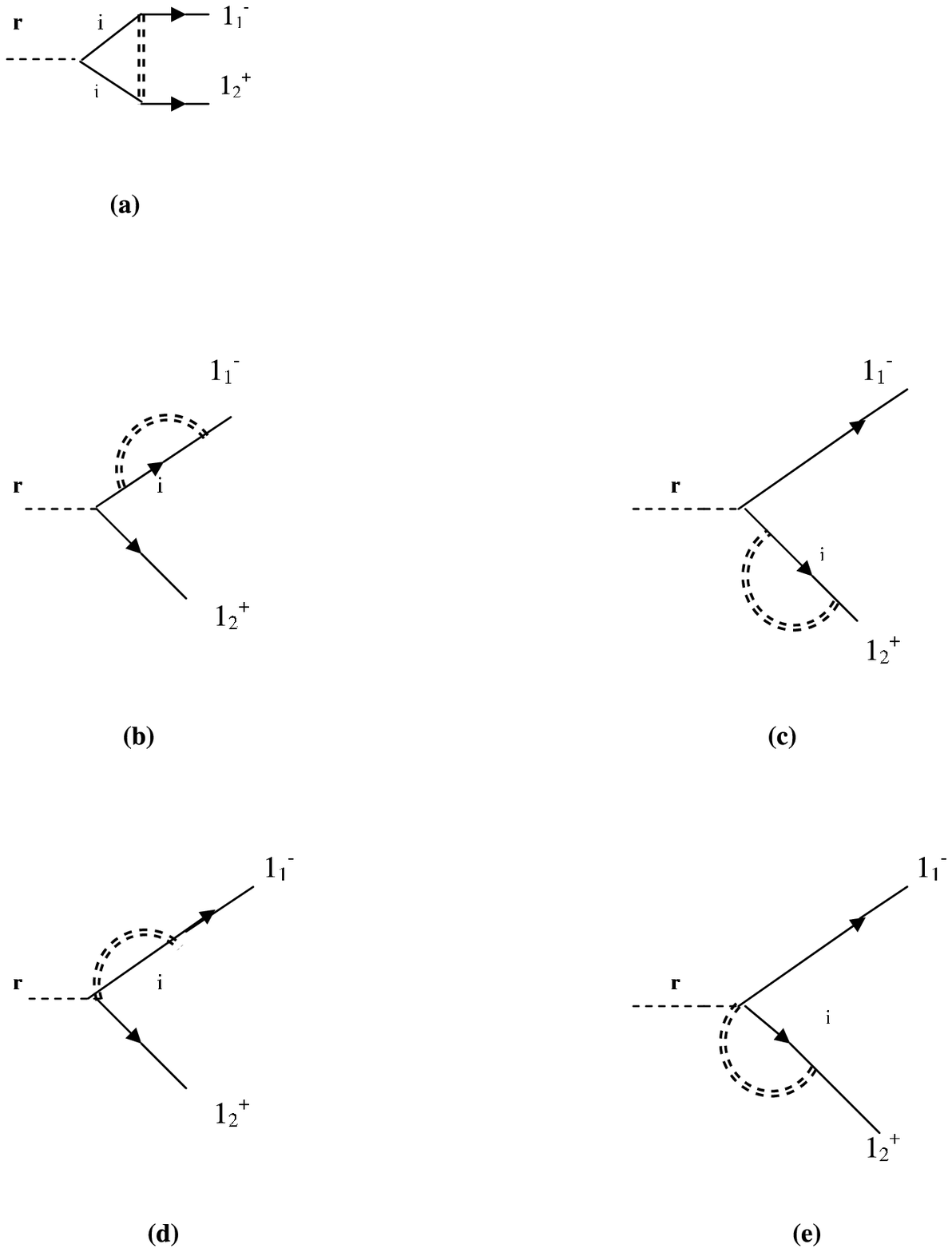} \vskip 3.5truein
\caption[]{One loop diagrams contribute to $r\rightarrow
l_1^-\,l_2^+$ decay with scalar unparticle mediator. Solid line
represents the lepton field: $i$ represents the internal lepton,
$l_1^-$ ($l_2^+$) outgoing lepton (anti lepton), dashed line the
radion field, double dashed line the unparticle field.}
\label{figselfvert}
\end{figure}
\newpage
\begin{figure}[htb]
\vskip -3.0truein \centering \epsfxsize=6.8in
\leavevmode\epsffile{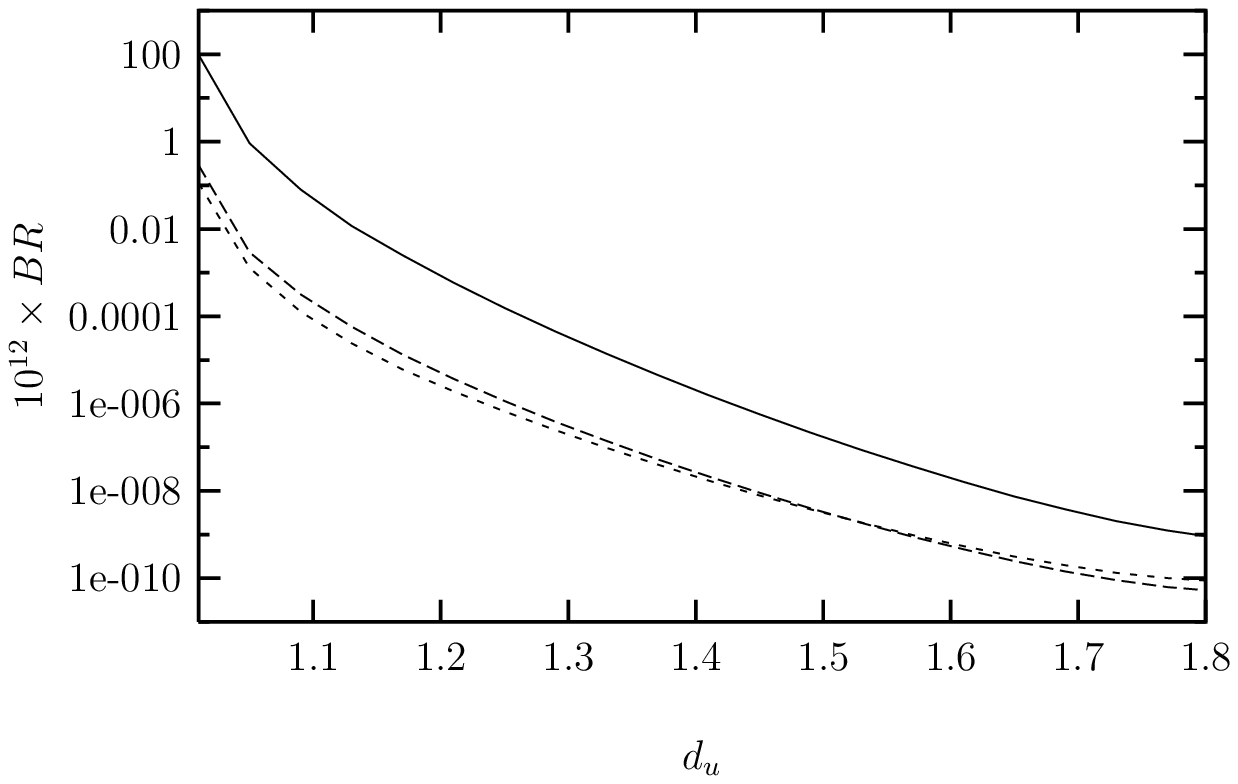} \vskip -3.0truein \caption[]{The
scale parameter $d_u$ dependence of the BR $(r\rightarrow
\mu^{\pm}\, e^{\pm})$, for $\Lambda_u=10\, (TeV)$, the couplings
$\lambda_{ee}=0.01$, $\lambda_{\mu\mu}=0.1$. The solid (dashed,
small dashed) line represents the BR for $m_r=200\,(GeV)$
($m_r=500\,(GeV),\, m_r=800\,(GeV)$).} \label{rmuedu}
\end{figure}
\begin{figure}[htb]
\vskip -3.0truein \centering \epsfxsize=6.8in
\leavevmode\epsffile{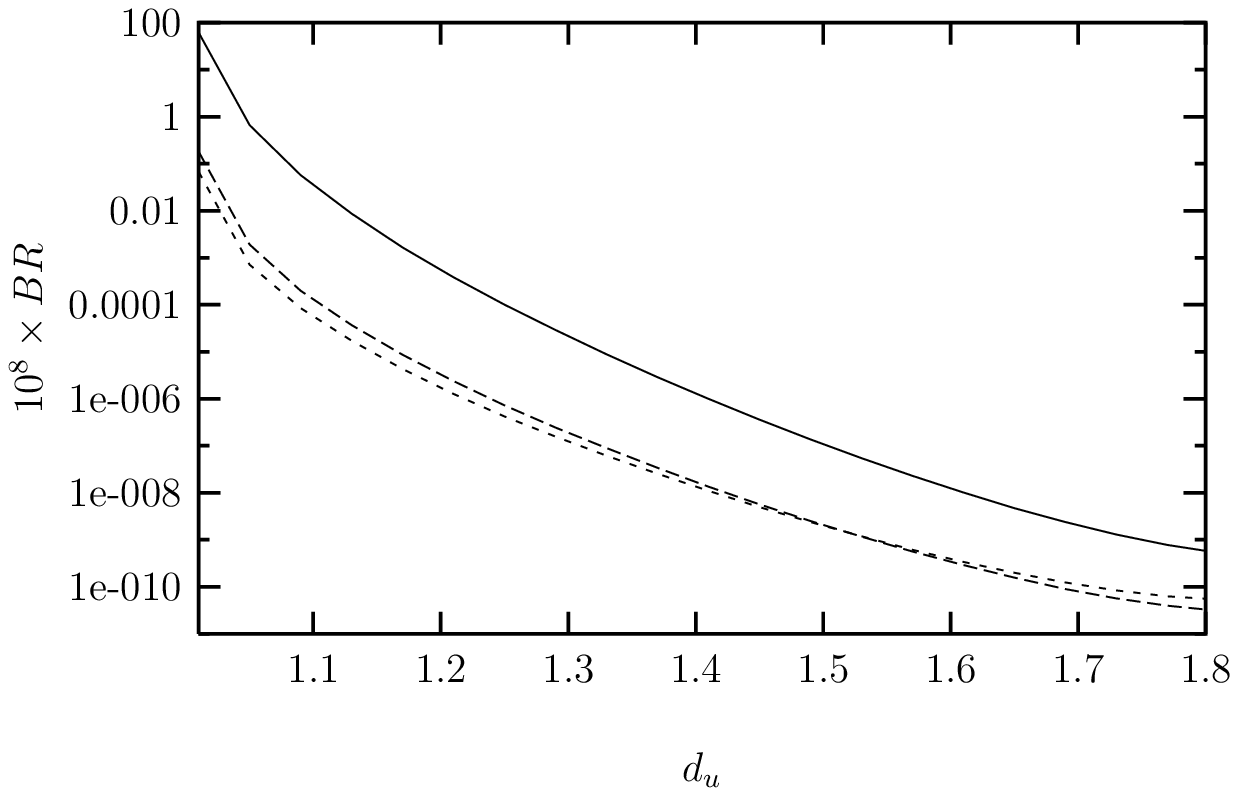} \vskip -3.0truein \caption[]{The
scale parameter $d_u$ dependence of the BR $(r\rightarrow
\tau^{\pm}\, e^{\pm})$, for $\Lambda_u=10\, (TeV)$, the couplings
$\lambda_{ee}=0.01$, $\lambda_{\tau\tau}=1.0$. The solid (dashed,
small dashed) line represents the BR  for $m_r=200\,(GeV)$
($m_r=500\,(GeV), m_r=800\,(GeV)$).} \label{rtauedu}
\end{figure}
\begin{figure}[htb]
\vskip -3.0truein \centering \epsfxsize=6.8in
\leavevmode\epsffile{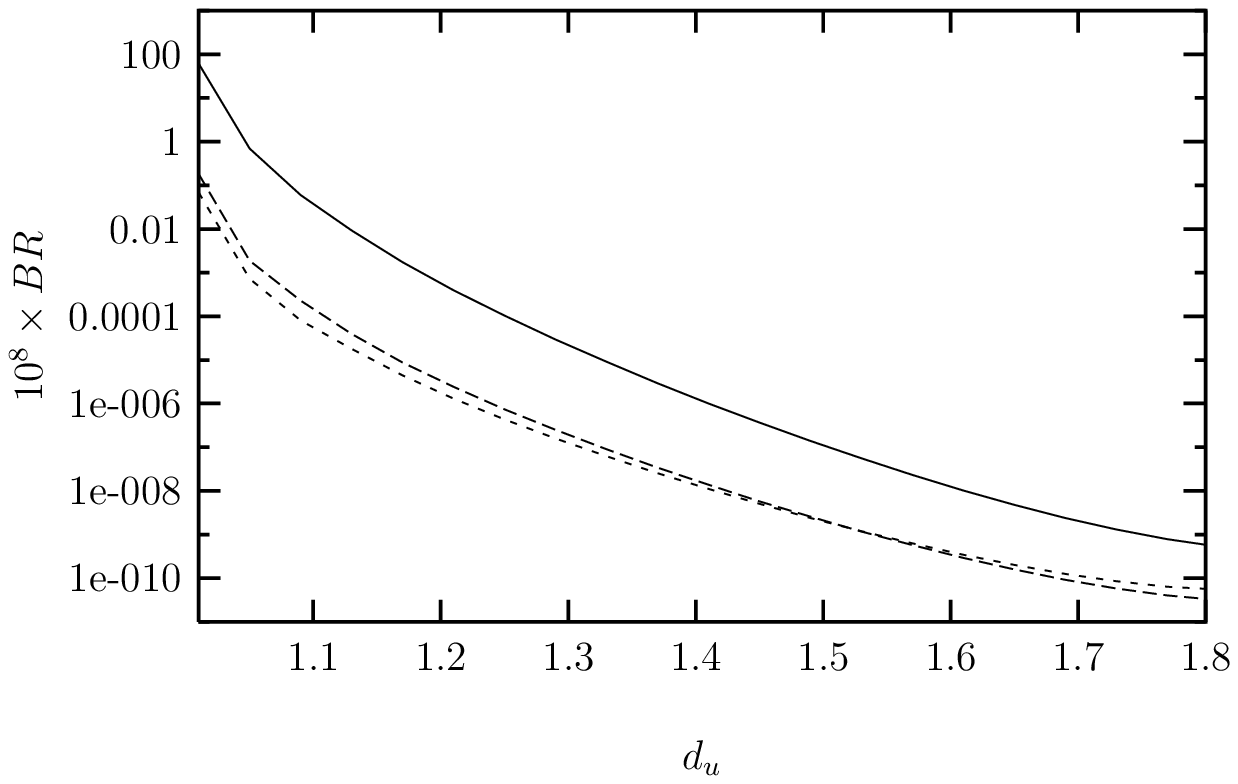} \vskip -3.0truein \caption[]{The
scale parameter $d_u$ dependence of the BR $(r\rightarrow
\tau^{\pm}\, \mu^{\pm})$, for $\Lambda_u=10\, (TeV)$, the
couplings $\lambda_{\mu\mu}=0.1$, $\lambda_{\tau\tau}=1.0$. The
solid (dashed, small dashed) line represents the BR  for
$m_r=200\,(GeV)$ ($m_r=500\,(GeV), m_r=800\,(GeV)$).}
\label{rtaumudu}
\end{figure}
\begin{figure}[htb]
\vskip -3.0truein \centering \epsfxsize=6.8in
\leavevmode\epsffile{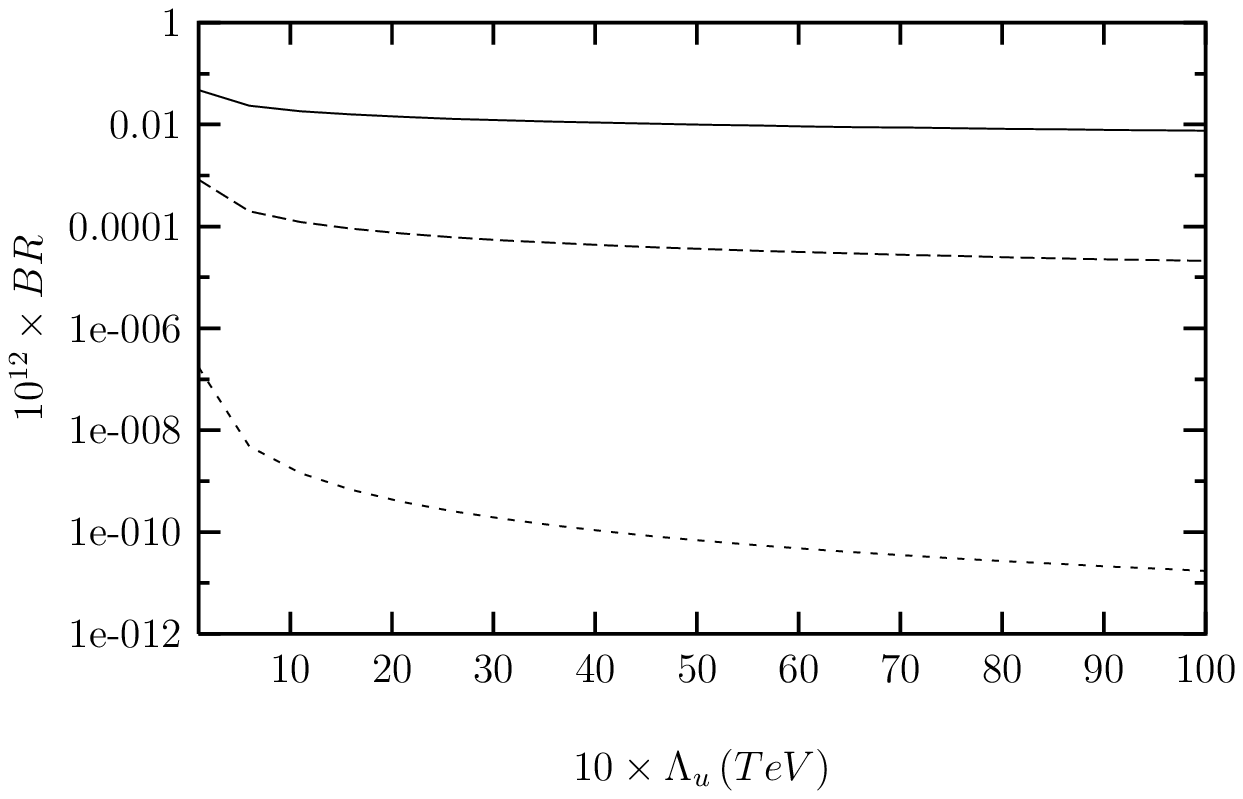} \vskip -3.0truein
\caption[]{$\Lambda_u$ dependence of the BR $(r\rightarrow
\mu^{\pm}\, e^{\pm})$, for $m_r=200\,(GeV)$, $\lambda_{ee}=0.01$
and $\lambda_{\mu\mu}=0.1$. Here the solid-dashed-small dashed
lines represent the BR for $d_u=1.1-1.2-1.5$.} \label{rmueLamu}
\end{figure}
\begin{figure}[htb]
\vskip -3.0truein \centering \epsfxsize=6.8in
\leavevmode\epsffile{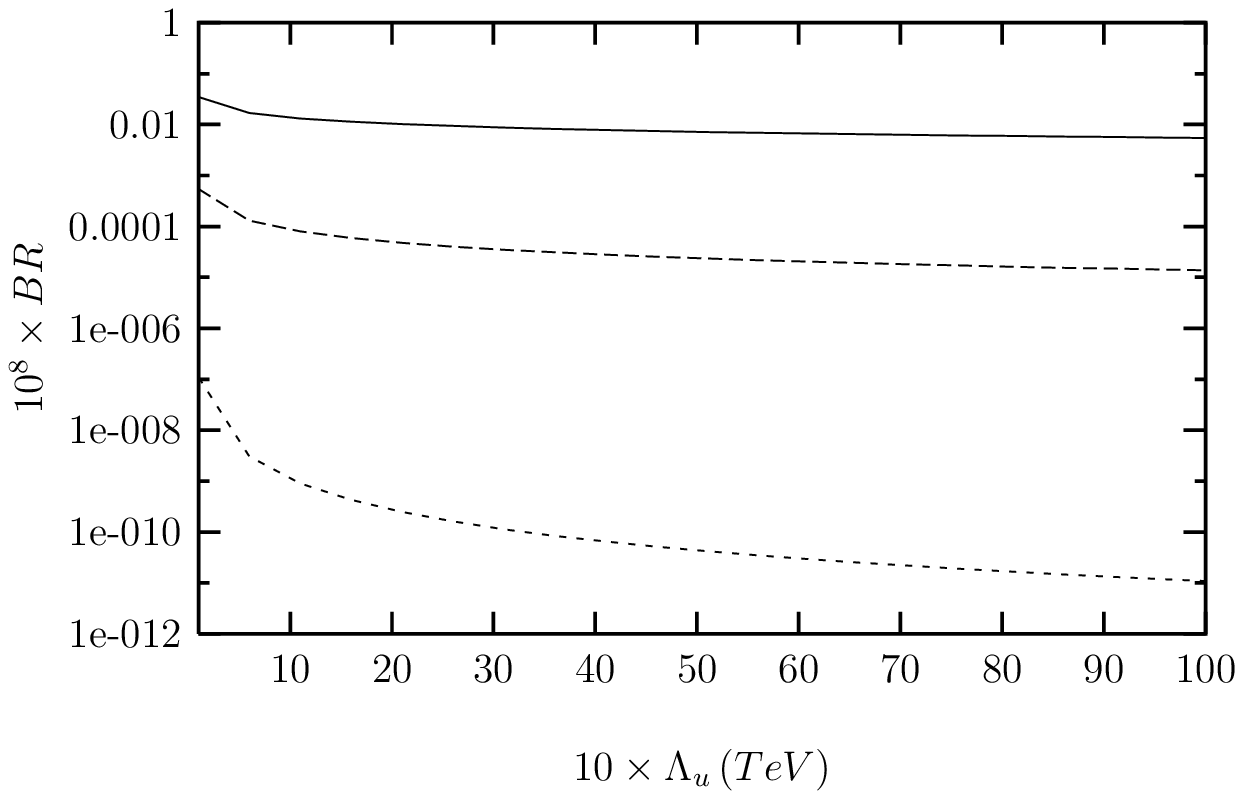} \vskip -3.0truein
\caption[]{$\Lambda_u$ dependence of the BR $(r\rightarrow
\tau^{\pm}\, e^{\pm})$, for $m_r=200\,(GeV)$, $\lambda_{ee}=0.01$
and $\lambda_{\tau\tau}=1.0$. Here the solid-dashed-small dashed
lines represent the BR for $d_u=1.1-1.2-1.5$.} \label{rtaueLamu}
\end{figure}
\begin{figure}[htb]
\vskip -3.0truein \centering \epsfxsize=6.8in
\leavevmode\epsffile{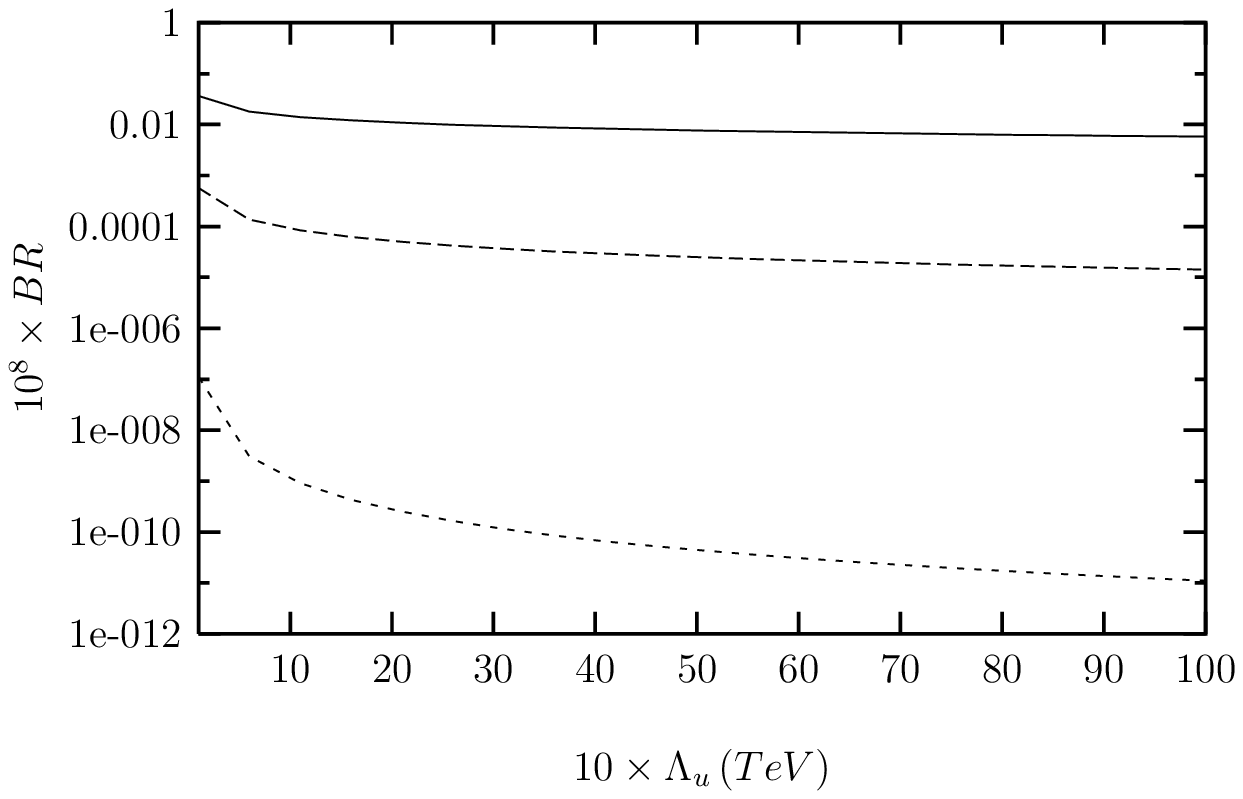} \vskip -3.0truein
\caption[]{$\Lambda_u$ dependence of the BR $(r\rightarrow
\tau^{\pm}\, \mu^{\pm})$, for $m_r=200\,(GeV)$,
$\lambda_{\mu\mu}=0.1$ and $\lambda_{\tau\tau}=1.0$. Here the
solid-dashed-small dashed lines represent the BR  for
$d_u=1.1-1.2-1.5$.} \label{rtaumuLamu}
\end{figure}
\begin{figure}[htb]
\vskip -3.0truein \centering \epsfxsize=6.8in
\leavevmode\epsffile{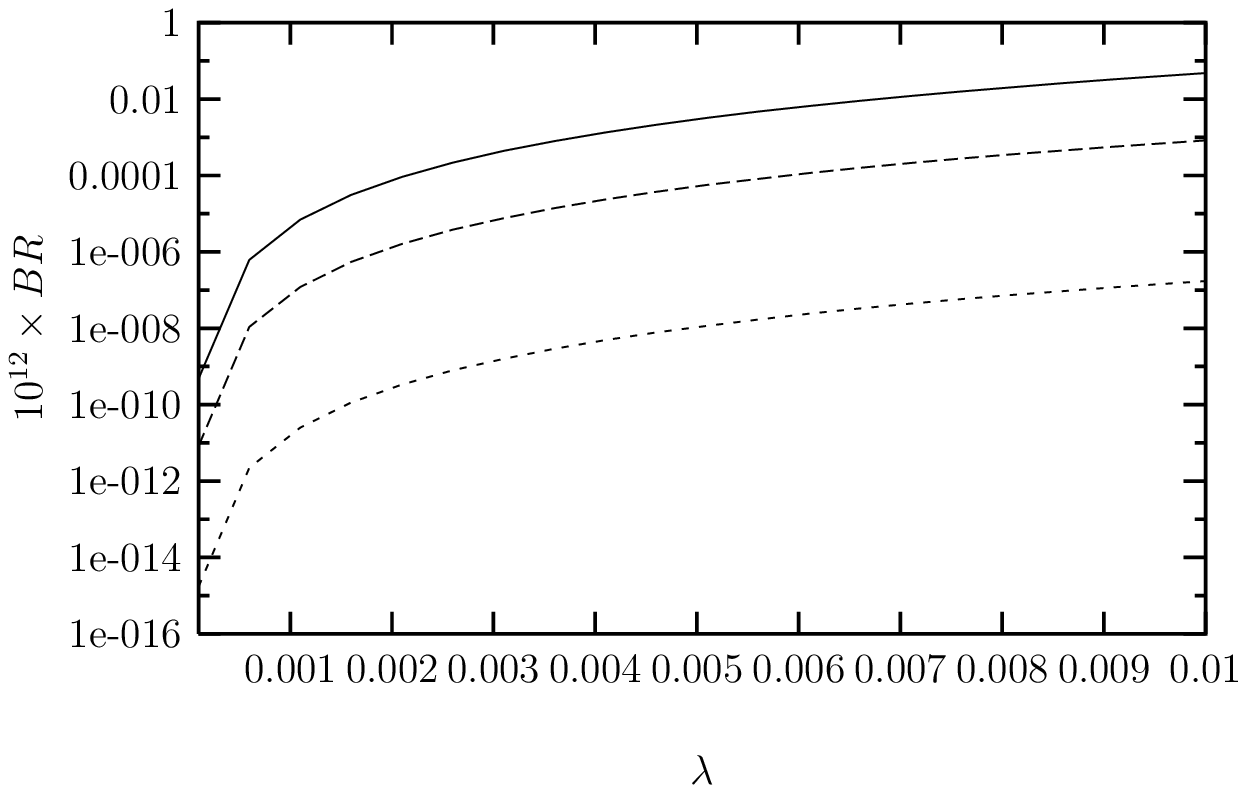} \vskip -3.0truein
\caption[]{$\lambda$ dependence of the BR $(r\rightarrow
\mu^{\pm}\, e^{\pm})$, for $m_r=200\,(GeV)$ and $\Lambda_u=10\,
(TeV)$. Here the solid-dashed-small dashed lines represent the BR
 for $d_u=1.1-1.2-1.5$.}
\label{rmuecoupl}
\end{figure}
\begin{figure}[htb]
\vskip -3.0truein \centering \epsfxsize=6.8in
\leavevmode\epsffile{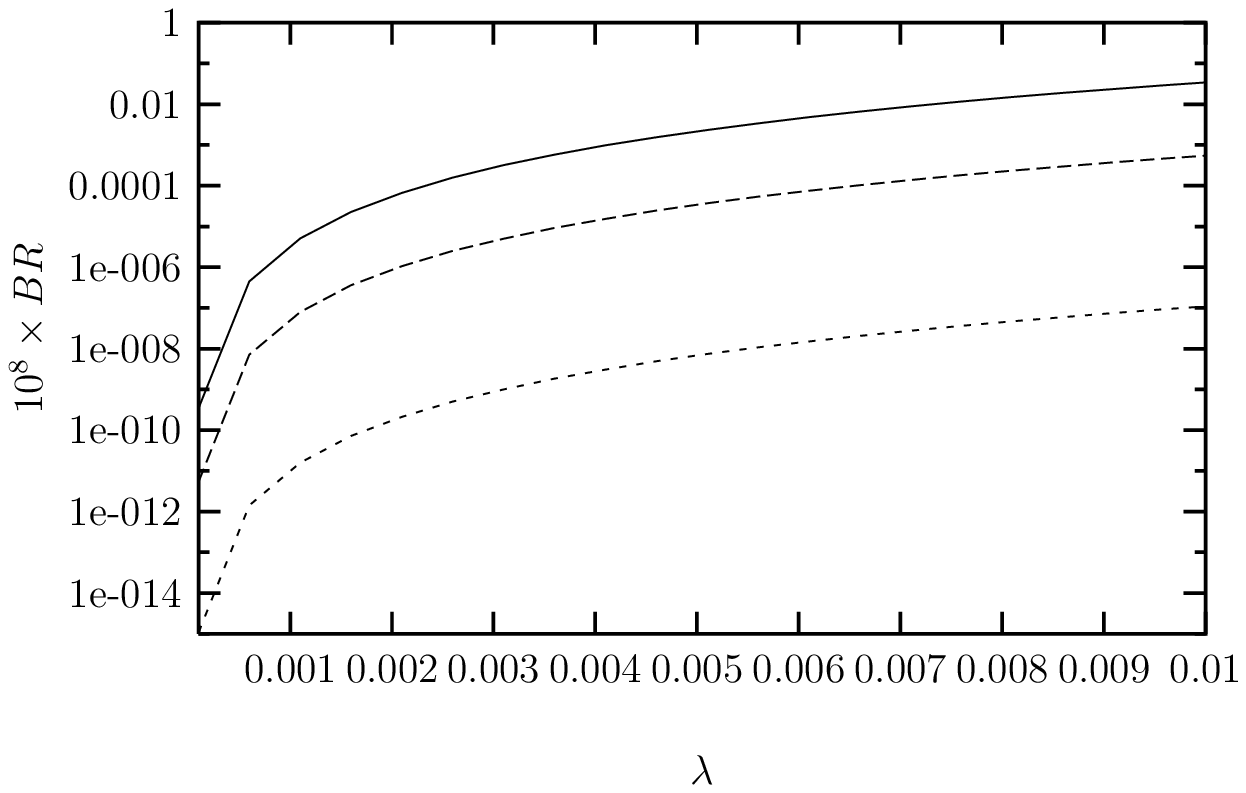} \vskip -3.0truein
\caption[]{$\lambda$ dependence of the BR $(r\rightarrow
\tau^{\pm}\, e^{\pm})$, for $m_r=200\,(GeV)$ and $\Lambda_u=10\,
(TeV)$. Here the solid-dashed-small dashed lines represent the BR
for $d_u=1.1-1.2-1.5$.} \label{rtauecoupl}
\end{figure}
\begin{figure}[htb]
\vskip -3.0truein \centering \epsfxsize=6.8in
\leavevmode\epsffile{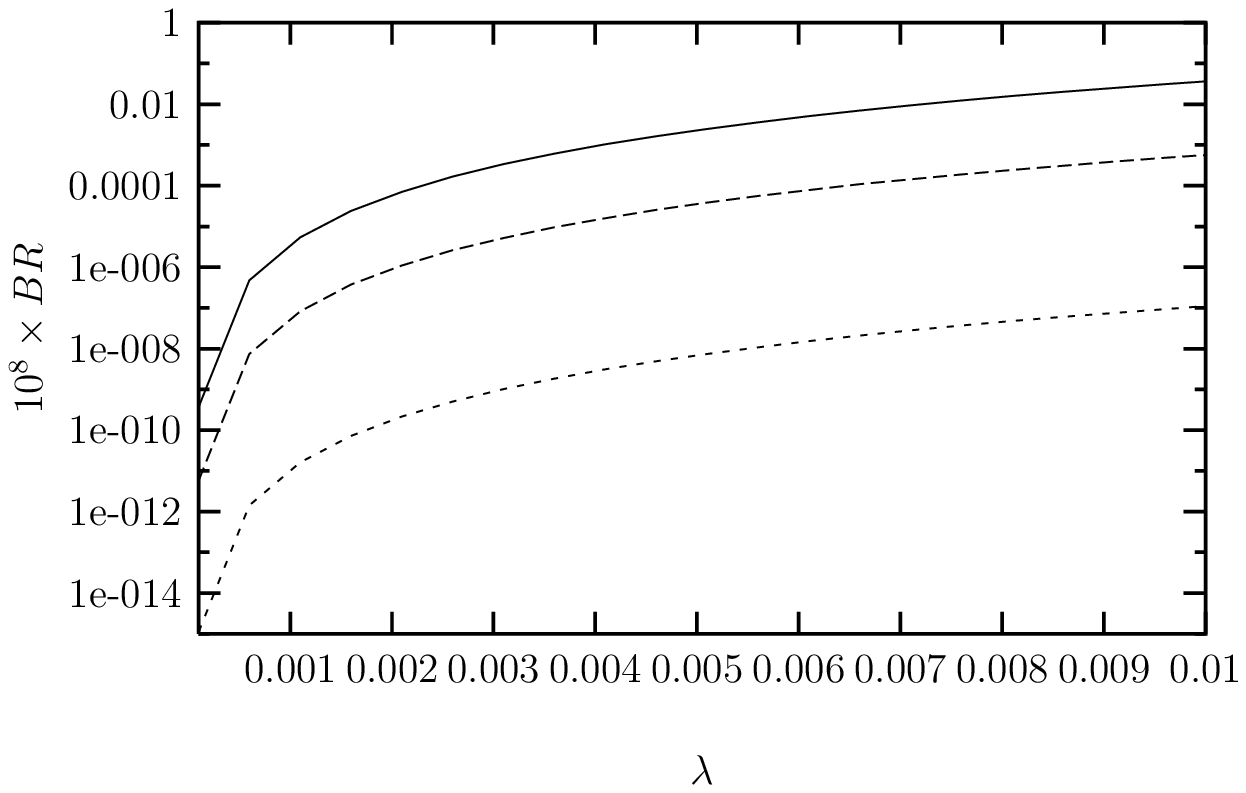} \vskip -3.0truein
\caption[]{$\lambda$ dependence of the BR $(r\rightarrow
\tau^{\pm}\, \mu^{\pm})$, for $m_r=200\,(GeV)$ and $\Lambda_u=10\,
(TeV)$. Here the solid-dashed-small dashed lines represent the BR
for $d_u=1.1-d_u=1.2-d_u=1.5$.} \label{rtaumucoupl}
\end{figure}
\end{document}